# Graph Data on the Web: extend the pivot, don't reinvent the wheel

*Fabien GANDON, Franck MICHEL, Olivier CORBY, Michel BUFFA, Andrea TETTAMANZI, Catherine FARON ZUCKER, Elena CABRIO, Serena VILLATA*
Inria, Université Côte d'Azur, CNRS, I3S, France, https://team.inria.fr/wimmics/

This article is a collective position paper from the Wimmics research team, expressing our vision of how Web graph data technologies should evolve in the future in order to ensure a high-level of interoperability between the many types of applications that produce and consume graph data. Wimmics stands for Web-Instrumented Man-Machine Interactions, Communities, and Semantics. We are a joint research team between INRIA Sophia Antipolis - Méditerranée and I3S (CNRS and Université Côte d'Azur). Our challenge is to bridge formal semantics and social semantics on the web. Our research areas are graph-oriented knowledge representation, reasoning and operationalization to model and support actors, actions and interactions in web-based epistemic communities. The application of our research is supporting and fostering interactions in online communities and management of their resources. In this position paper, we emphasize the need to extend the semantic Web standard stack to address and fulfill new graph data needs, as well as the importance of remaining compatible with existing recommendations, in particular the RDF stack, to avoid the painful duplication of models, languages, frameworks, etc. The following sections group motivations for different directions of work and collect reasons for the creation of a working group on RDF 2.0 and other recommendations of the RDF family.

## Make Graph Data on the Web *the keystone*

There is now a long tradition of graph-based data and knowledge representation. Graphs, networks, relational or linked data are today found everywhere (e.g. social media, connected devices / IoT, transportation/energy/telcom networks, biological pathways, etc.). But before discussing new graph languages, models or extensions, we found it important to recall that there exists a long history of graph-based semantic representations and an extensive literature on them.

Semantic networks have a long history in philosophy, psychology and linguistics [7], starting long before the birth of computer science. They provide graphic notations for modeling graph data and representing knowledge ranging from highly informal to logically grounded and formally defined formalisms. For instance, in his "history of semantic networks" Sowa mentions [8] the tree notation for first-order logic by Gottlob Frege (1879), the graphic notation for predicate calculus and the relational graphs of Charles Sanders Peirce (1882), the dependency graphs of Lucien Tesnière (1959), the correlational nets of Silvio Ceccato (1961), the conceptual dependency graph of Schank & Tesler (1969), propositional semantic network in MIND by Stuart Shapiro (1971), conceptual graphs by Sowa (1976), etc. Several families of semantic networks are even identified by [7] including definitional networks, assertional networks, implicational networks, executable networks, learning networks and hybrid networks combine two or more of these families.

This reminder of the number and variety of graph-inspired languages and formats that were proposed is mainly to insist on the necessity to stop adding new languages and start evolving existing ones, in particular the RDF pivot language. One of the achievements of RDF was to propose a unified and universal model for graph data representation. It is now a standard model used world-wide on the Web in public applications (e.g. library data publication, rich snippets, open biological databases, HTML pages markup) and private applications (e.g. data integration, pivot format between legacy systems, intranets/extranets). RDF managed to standardize a graph data model and knowledge representation at the World-Wide scale of the Web, enabling the production and consumption of knowledge from heterogeneous systems (graph bases, sensors, prolog-based systems, relational and NoSQL applications, etc.) This achievement should be consolidated by maintaining, extending and evolving the RDF galaxy rather than proposing alternative graph formalisms at the risk of duplicating the entire stack of standards (model, syntaxes, validation, querying, schemas, etc.) and losing the benefit of a unified and universal model for graph data representation supporting a maximum of the graph data variety (graph data streams, geo-spatial graph data, temporal graph data, etc.)

As a concrete example, today, it has become a commonplace to highlight, to say the least, the differences between two closely related data graph approaches, namely property graphs and RDF graphs. Possibly fueled by parochial conflicts more than objective conceptual and technical discrepancies, this apparently irrevocable distinction seems somehow artificial though. Indeed, both approaches model directed graphs wherein nodes and edges are labeled. A difference that is often excessively stressed is with respect to the management of properties: while they are attached to nodes as (key, value) pairs in property graphs, RDF models them as regular edges whose targets are literal values. Yet, this difference is more of an implementation choice (that may have significant impacts on performance) than a major conceptual discrepancy. Indeed, an RDF store implementor may choose to treat edges differently depending on the kind of their target (URI and blank node vs. literal), for instance to set up specific indexes of literal values allowing for free text search. A second difference, surprisingly less stressed than the earlier although it is probably more crucial, concerns the way one can attach properties to edges. Property graphs manage such properties exactly as in the case of nodes, as (key, value) pairs. In RDF, this process called reification allows to write triples about triples. This mechanism has proved not to be efficient insofar as (i) it requires the addition of multiple triples and (ii) it is not managed transparently in SPARQL queries: one has to denote explicitly the reified triples that the query is meant to match. Several alternatives have been proposed [3], such as extensions of RDF, SPARQL and Turtle [4] that are promising leads.

But the key point for us here is that the discrepancies between RDF and property graphs, when justified by motivating scenarios, should be considered not as an opposition where one should win over the other eventually, but as requirements for an RDF 2.0 specification that maintains a unified and universal model. RDF 2.0 should preferably be backward compatible with RDF 1.1 and the stack of standards above in order to avoid the duplication of models, languages, frameworks and the inevitable next step of mapping and translation between the two worlds. RDF was positioned as the graph data model of the Web architecture and we should strive to maintain it as such to lead the Web to its full potential of interoperability platform.

# Make Graph Data on the Web *interoperable*

As explained in the previous section, we believe that the evolution of RDF (its model and syntaxes) should emphasize the importance of maintaining and confirming it as a pivot language for graph data interoperability. Use cases considered for the evolution of RDF should span needs from entreprise knowledge graphs, Web of Things, property graphs, etc. In particular, below we identify some of the issues or limitations that, we believe, should be addressed in a global reflection about the evolution of RDF, to ensure its suitability as a universal interoperability format applicable to a large scope of scenarios.

The fact that RDF triples are binary relations can be a limitation. There are several use cases where we would like to have additional arguments: temporal information, modality, provenance, value with uncertainty, value with unit, not speaking of natively n-ary relations. These needs are also related to the ability to assign a unique URI to a triple in such a way that we can associate metadata to triples without resolving to the reification or to the multiplication of named graphs. The ability to use triples as the subject or object of other triples is both needed to support all the scenarios requiring metadata about the triples (e.g. time stamp, provenance) and to improve the ability of RDF to play its role of pivot language in the portability of graph data across systems.

In other scenarios, we would like to have an order on edges of a given node, like in an XML tree. That is, we would like to distinguish the first edge, the second edge, etc. even if the properties are of different types.

The introduction of named graphs in RDF 1.1 has spawn the emergence of new needs such as the ability to nest graphs and more generally the identification of standard relation types between named graphs, such a special properties to specify an entailment regime for a specific named graph.

Another issue pertains to RDF's lack of a native data structure for managing collections. An RDF list (rdf:List) consists of a head (rdf:first), a rest represented as another list (rdf:rest) and the special end-of-list marker (rdf:nil). Other collection data types such as rdf:Bag, rdf:Seq and rdf:Alt rely on an additional unlimited set of predicates rdf:_1, rdf:_2, etc, to name each of the collection members. Such complex structures make managing collections extremely cumbersome and developer-unfriendly. As an consequence, for instance, the LDScript SPARQL extension language [1] introduces a list data type that allows manipulating lists of RDF terms more naturally like in most programming languages. To aim for a generic pivot language, RDF needs new native data types to represent collections as first-class citizens, not as complex assembly of triples.

Finally the Web interoperability was also leveraged for many internal applications of our organizations. Many of our intranets are, in fact, largely intrawebs and the same could be said about extranets linking our digital ecosystems. The data and application integration inside a lot of companies now rely on internal linked data, triple stores and knowledge graphs requiring support for schema alignment (generation and representation) and resource alignment (named entity linking representation, detection, collection). In addition, and again, a key point is to encourage the adoption of RDF and SPARQL as standards for the graph databases just like SQL and the relational model in classical databases.

# Make Graph Data on the Web *developer friendly*

A key feature of the adoption of any Web standard is its degree of developer friendliness. Adoption relies on the availability of tools and the support for developers to use them, integrate them in existing languages and systems and deploy them. Developing on top of linked data could be made easier with a number of recommendations improving the life of programmers:

- Specify API for Web graph data (just like the DOM) to be reused across languages, across implementations and libraries. The lack of standard RDF API makes it hard to switch between implementations, distribute tools or extensions beyond one specific framework, combine several solutions in one system, etc. Moreover each API requires a learning curve from the developers and ad hoc wiring to be integrated. A standard JavaScript API for RDF could be embedded in Web Browsers, making CRUD operations on RDF data easier to perform from JavaScript, i.e for manipulating and displaying SPARQL results. Front-end developers should be able to work with RDF data as objects, not as string based triplets. A first initiative ([http://rdf.js.org/](http://rdf.js.org/) ) for a low level interface definition representing RDF data independently of a serialized format in a JavaScript environment has been initiated by the [RDF JavaScript Libraries Community Group](). It could be a good starting point for a standard RDF API in the browser.
- Programming languages with native Web graph data structure would help the rapid prototyping and perennial development. When RDF becomes a native datatype of the language and when the language natively provides high level operators for RDF, it becomes easier to abstract algorithms from the syntax and related low-level concerns such as parsing data structures, syntactic sugar and variations, basic access and manipulation operations, etc.
- HyperGraphQL ([http://hypergraphql.org/](http://hypergraphql.org/)) or GraphQL-LD, are GraphQL-based interfaces for querying RDF stores. Their goal is to simplify the production and consumption of RDF data through SPARQL endpoints for use by other APIs or frameworks (ex: React, Angular). The proposed model consists of transforming GraphQL queries coupled with a JSON--LD context into SPARQL queries, and of converting SPARQL results to a GraphQL-query -compatible response (a JavaScript simple object). Such an approach could be considered to propose several levels of complexity when a front-end developer has to deal with RDF data, depending on the applications that will be developed: 1) a low-level approach that would be close to the abstraction level of the DOM API, and a 2) higher-level approach that would propose concepts similar to what HyperGraphQL offers (see [15] and [16] for related works).

# Make Graph Data on the Web *secure and trusted*

A number of use cases for standard graph data exchange on the Web require the ability to ensure the integrity of a graph content and the authenticity of a graph producer or consumer. For instance, linked data stating that a person graduated from a given university along with the identity of the issuing institution need to be certified to be trusted and used by applications.

More generally, any scenario requiring authentication, secured graph data publication, verifiable claims, non-repudiation, etc. is hardly achievable today due to the lack of a standard canonical RDF graph representation which is a necessary step to support signing

RDF graphs. Therefore we believe there is a need to start a recommendation stack above RDF with two first layers: a recommendation for RDF graphs canonicalization, and a recommendation for signing RDF graphs. Relevant works in that direction include [9][10] and [11].

In addition to these first steps, we can identify the challenges of distributed and secured access control to graph data including recommended practices for secured and controlled dereferenciation. Security also creates new situations such as the existence of an HTTPS and an HTTP URIs to refer to the same resource and the possibility to have different answers for each of them while raising the question of their relations.

# Make Graph Data on the Web *viewable*

Since the initial initiative of FRESNEL [12] to provide a display vocabulary for RDF, and apart from some extensions such as PRISMA for context-aware adaptation for linked data [13], the problem of presenting linked data in a human-readable manner has not been considered at W3C. Other initiatives include the work on RDF-dedicated browsers relying on the linked data principles to support browsing linked data in a generic and domain independent way such as the Tabulator [14].

The more graph data we publish on the Web, the more we need to provide application-independent standards to attach visualization and interaction metadata to RDF graphs in order to support their display in a user-friendly way. FRESNEL had that ambition to provide a selection mechanism and formatting language to allow us to specify which piece of data should be displayed and how. Relying on the current standards (SPARQL, SHACL, CSS) we could reconsider this pioneer initiative and provide a standard to publish RDF graph style sheets that could play the same role than CSS but for RDF data rendering purpose. It would be a big asset in bridging the Web of Data and the classical Web applications and pages.

# Make Graph Data on the Web *compatible with uncertainty*

Current standards of the Semantic Web and Linked Data do not support the representation of incomplete, uncertain, or inconsistent data graphs. Uncertainty may take the form of inconsistencies, ambiguity, fuzziness, incompleteness, etc. But in all its forms, uncertainty becomes an obstacle to data reuse as soon as it is not made explicit and available with the data: in many scenarios, an external source cannot be used if the reliability of its data cannot be assessed. Several proposals have been made to extend the current standards [6] and we believe this is an important extension to augment the adoption and use of graph data found on the Web. Moreover, after the representation of uncertainty comes the question of its processing (e.g. approximate reasoning, fuzzy reasoning) and the current semantic Web framework does not support these types of reasoning without extensions to its recommendations.

## Make Graph Data on the Web *distributed in storage and processing*

Interconnecting distributed graph data available from multiple providers is a requirement in many use cases and applications today. Unlike graph databases that are often implemented as closed worlds similar to relational databases, the Linked Data principles provide a grounding to deploy distributed graph data. One mechanism for querying linked data is to dereference URIs to access representations of resources, preferably in RDF, whose content refers to other URIs that, in turn, can be dereferenced. This enables applications to seamlessly exploit distributed data sources using a follow-your-nose navigation approach. Alternatively, a SPARQL endpoint can execute queries against a local triple store or federated queries distributed over different SPARQL endpoints. Yet, several issues still hamper the seamless querying of distributed RDF data sources. Below we identify a few of them.

Although the Linked Data principles recommend that looking up (dereferencing) URIs should return informative RDF documents, there is no way to know in advance whether a URI is dereferenceable. Future evolutions of the semantic Web stack may address this lack by clarifying the semantics of dereferenceable URIs. Existing works may be leveraged, such as Hydra [5] that represents in a machine-readable manner if a URI is intended to be dereferenced or solely considered as an identifier, or other methods (e.g. heuristics, machine learning) to assess whether a URI can be dereferenced to RDF content.

As of today, there is a strict separation between Linked Data navigation and SPARQL query processing: with the exception of the SPARQL "FROM NAMED" clause that, under some interpretation, can suggest downloading the content of a named graph, a SPARQL endpoint does not dereference URIs to retrieve additional data upon which the query could be processed. The SERVICE clause gives the URL of a SPARQL endpoint against which a subquery is to be evaluated, whereas the GRAPH clause names the URI of a graph hosted in the local triple store. In a generalized approach, the SERVICE clause could allow querying not only third-party SPARQL endpoints but also any source providing RDF content: an RDF file in any RDF serialization syntax, a named graph URI that dereferences to the RDF content of the graph, a Web page containing markup data embedded as RDFa or JSON-LD etc., as proposed in SPARQL-LD [2].

Finally, clear semantics would be needed to process distributed SPARQL queries on federated SPARQL endpoints, specifying for example how to treat distributed named graphs, how to deal with blank nodes, duplicate results due to duplicate triples, Property Path, etc. Named graph URI dereferencing deserves a precise semantics that is still to be formalized. In addition, SPARQL could be extended in order to specify the list of SPARQL endpoints URL to be queried to solve a distributed query.

## Make Graph Data on the Web *a distributed AI enabler*

Data is fueling the latest AI advances and clearly became an extremely important resource to train and apply a variety AI techniques simulating different intelligent behaviors (learning, reasoning, identifying, interacting, etc.) each of them coming in many forms, for instance for reasoning: deductive, inductive, abductive, fuzzy, temporal, causal, etc

But RDF and Linked data can provide much more than input data. They provide schemas to label data, enrich them and train from them. They can be used to represent and publish output data, to attach metadata (provenance, precision, algorithms used, detected bias), to publish and reuse parameters, embeddings, configurations and even entire networks. RDF and linked data can be used for and by AI at many more levels than just input data.

The special case of schemata is especially important since they provide semantics useful in very different processing (e.g. reasoning and learning both use classes). Recommendations and infrastructures are needed to support the life-cycle of vocabularies and encourage their re-use across applications fostering, again, interoperability. Recommendations include the provision of standard vocabulary metadata to facilitate their discovery and reuse. Key services include schemata search engines and directories (e.g. LOV) and platforms natively supporting low level aspects of schema publishing (dereferenceability, content negotiation and documentation, redirect, multiple format/syntax support, etc.)

The current standards of the W3C supporting some types of reasoning could also be improved. For instance the relation between RIF and RDF is not optimal to support rule-based reasoning and many systems prefer to use SPARQL and its syntax to express rules raising the question of having a SPARQL based/inspired recommendation for rule-based system in RDF.

In the distributed context of the Web, RDF as the potential to become an AI lingua franca for graph data and semantic networks and to support distributed AI architectures (including the emerginging edge AI at W3C) and system combining or hybridizing multiple AI methods. In addition, the growing complexity of these systems requires to support explainable AI approaches producing interpretable, explainable and justified results. The combination of hypermedia HATEOAS (Hypermedia as the Engine of Application State) and linked data approaches can provide a universal framework to produce, publish and access the traces of the complex AI systems providing and processing data.

# Conclusion: Maintain RDF as the Graph Data standard on the Web

In this collective position paper we collected several examples of extension of the semantic Web standard stack to address and fulfill new graph data needs. We also stressed the importance of ensuring the compatible with existing recommendations and, in particular, in extending and build on top of the RDF galaxy. This, in our opinion, is vital to lead the Web of its full potential in linking all forms of intelligence.

///
////

query languages for graph databases

approaches for transforming data between different vocabularies with overlapping semantics,
what is needed to better support work on vocabulary standards.

relational databases use the standard SQL for query and update
SQL databases that address the need for flexible handling of unstructured data with key-value stores, document stores, and graph databases.
graph databases come into their own, as they enable links across data, for fine grained networks of information. Graph databases may support robust transactions
BUT they need to be playing the game of Standards and fully compatible with RDF and SPARQL in particular.

in the use of graph databases, time-series databases, geo-spatial data, temporal data implications for query languages for graph data that incorporates indexing for temporal and spatial annotations

Paving the WAI article.
Foundations of an Alternative Approach to Reification in RDF
https://arxiv.org/abs/1406.3399

Federation, distribution, more/less intelligent servers/clients

Motivations for a WG on RDF 2.0 for a paper at:
https://www.w3.org/Data/events/data-ws-2019/index.html

History of Semantic Networks
http://www.jfsowa.com/pubs/semnet.htm
Quillian, Sowa,

1971 The first propositional semantic network to be implemented in AI was the MIND system, developed by Stuart Shapiro (1971)
Woods (1975) and Brachman (1979) DL and KL-ONE
“Conceptual graphs (Sowa 1976,) are a variety of propositional semantic networks in which the relations are nested inside the propositional nodes. They evolved as a combination of the linguistic features of Tesnière’s dependency graphs and the logical features of Peirce’s existential graphs with strong influences from the work in artificial intelligence and computational linguistics”
WebKB, P. Martin 199x CGs on the Web
RDF

<add schema from slides> <+ Dagstuhl summary and Ref>